\documentclass[aps,prl,twocolumn,showpacs,groupedaddress]{revtex4-1}  
\usepackage{graphicx, amsmath}  

\begin{document}


\title{Nonlinear excitation of acoustic modes by large amplitude Alfv{\'e}n waves in a laboratory plasma}


\author{S.~Dorfman}
\author{T.~A.~Carter}
\affiliation{University of California Los Angeles, Los Angeles, California 90095, USA}
\vskip 0.25cm


\date{\today}

\begin{abstract}
The nonlinear three-wave interaction process at the heart of the parametric decay process is studied by launching counter-propagating Alfv{\'e}n waves from antennas placed at either end of the Large Plasma Device (LAPD).  A resonance in the beat wave response produced by the two launched Alfv{\'e}n waves is observed and is identified as a damped ion acoustic mode based on the measured dispersion relation.  Other properties of the interaction including the spatial profile of the beat mode and response amplitude are also consistent with theoretical predictions for a three-wave interaction driven by a non-linear pondermotive force.
\end{abstract}


\maketitle


Alfv{\'e}n waves, a fundamental mode of magnetized plasmas, are
ubiquitous in lab and space.  While the linear behavior of these waves
has been extensively studied \citep{alfven42,wilcox60,morales97,gekelman99,vincena04}, non-linear effects
are important in many real systems, including the solar wind and solar
corona.  In particular, a parametric decay process in which a large
amplitude Alfv{\'e}n wave decays into an ion acoustic wave and
backward propagating Alfv{\'e}n wave may play a key role in
establishing the spectrum of solar wind turbulence \citep{zanna01}.
  Ion acoustic waves have been observed in the heliosphere, but their
  origin and role have not yet been determined \citep{mangeney99}.
  Such waves produced by parametric decay in the corona could
  contribute to coronal heating \citep{pruneti97}.  Parametric decay
  has also been suggested as an intermediate instability mediating the
  observed turbulent cascade of Alfv{\'e}n waves to small spatial
  scales \citep{zanna01,yoon08}.

In this letter, the first laboratory observations of the Alfv{\'e}n-acoustic mode coupling at the heart of the parametric decay instability are presented.  Counter-propagating shear Alfv\'{e}n waves are launched from antennas and allowed to interact nonlinearly.  As the beat frequency between these two launched waves is varied between discharges, a resonant response is observed when frequency and wave number matching is satisfied for coupling to an ion acoustic mode.   Other features of the interaction including the beat mode spatial structure and response amplitude match predictions based on a three-wave interaction driven by a non-linear pondermotive force.

Although these results represent a beat wave process rather than an instability, the reported evidence of a three-wave interaction may be used to validate simple theoretical predictions and aid in comparison with space measurements.  To date, there has been an abundance of theoretical work \citep{sagdeev69,hasegawa76,goldstein78,wong86,longtin86,hollweg93,hollweg94}, but very little direct experimental observation of parametric decay.  Observations by \citet{spangler97} in the ion foreshock region upstream of the bow shock in the Earth's magnetosphere indicate the presence of large amplitude Alfv{\'e}n waves as well as density fluctuations with no magnetic spectral component; the latter are presumed to be acoustic modes resulting from parametric decay, but detailed measurements are limited.

\begin{figure}[tbp]
\centering
\includegraphics[width=\columnwidth]{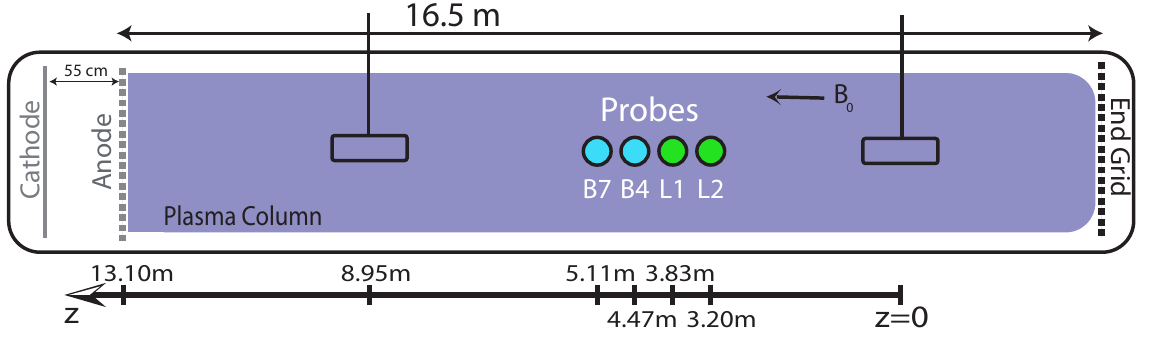}
\caption{Experimental setup in the LAPD plasma column.  Alfv{\'e}n wave antennas shown at either end of the device launch the counter-propagating Alfv{\'e}n waves to be examined in this study.  Magnetic probes B4 and B7 and Langmuir probes L1 and L2 placed between the two antennas in the plasma column are used to diagnose the interaction; various $z$ positions are used for the data in the paper.}\label{fig:esetup}
\end{figure}

The Large Plasma Device (LAPD) at UCLA is an ideal environment for
experiments diagnosing nonlinear Alfv\'{e}n wave interactions.  The LAPD is a cylindrical
vessel capable of producing a $16.5$~m long, quiescent, magnetized plasma
column for wave studies.  The BaO
cathode discharge lasts for $\sim{10}$~ms, including a several ms
current flattop.  Typical plasma parameters for the present study are
$n_e\sim10^{12}$~/cm$^3$, $T_e\sim5$~eV, and $B_0\sim400-900$~G
($\beta\ll1$) with a fill gas of helium or hydrogen.  Extensive
prior work has focused on the properties of linear Alfv{\'e}n waves
\citep{leneman99,gekelman00,vincena04,palmer05}.  Studies of the
nonlinear properties of Alfv\'{e}n waves have also been performed on
LAPD; in these experiments, two co-propagating Alfv{\'e}n waves non-resonantly drive a quasi-mode in the plasma \citep{carter06} or resonantly drive drift wave instabilities \citep{auerbach10}.

For the present set of experiments, loop antennas placed at either end
of the LAPD, shown in Fig.~\ref{fig:esetup} launch linearly polarized,
counter-propagating Alfv{\'e}n waves with amplitudes of
$\delta{B}\sim1$~G during the discharge-current-flattop period of the LAPD discharge.  It should be noted that parametric decay of a single Alfv{\'e}n wave is not observed in these experiments; consistent with this, the experimental value of $\delta{B}/B_0\leq2\times10^{-3}$ gives a growth rate \citep{sagdeev69} that is comparable to an Alfv{\'e}n wave transit time through the entire plasma column for relevant experimental parameters.  Instead, antennas directly launch both the ``pump'' and ``daughter'' Alfv\'{e}n waves at similar amplitudes.  In the plasma column between the antennas, magnetic probes detect the magnetic field signatures of the launched modes while Langmuir probes are used to detect signatures of a density response at the beat frequency.  Each probe is mounted on an automated positioning system that may be used to construct a 2-D profile in the $x$-$y$ plane averaged across multiple discharges.

\begin{figure}[tbp]
\centering
\includegraphics[width=\columnwidth]{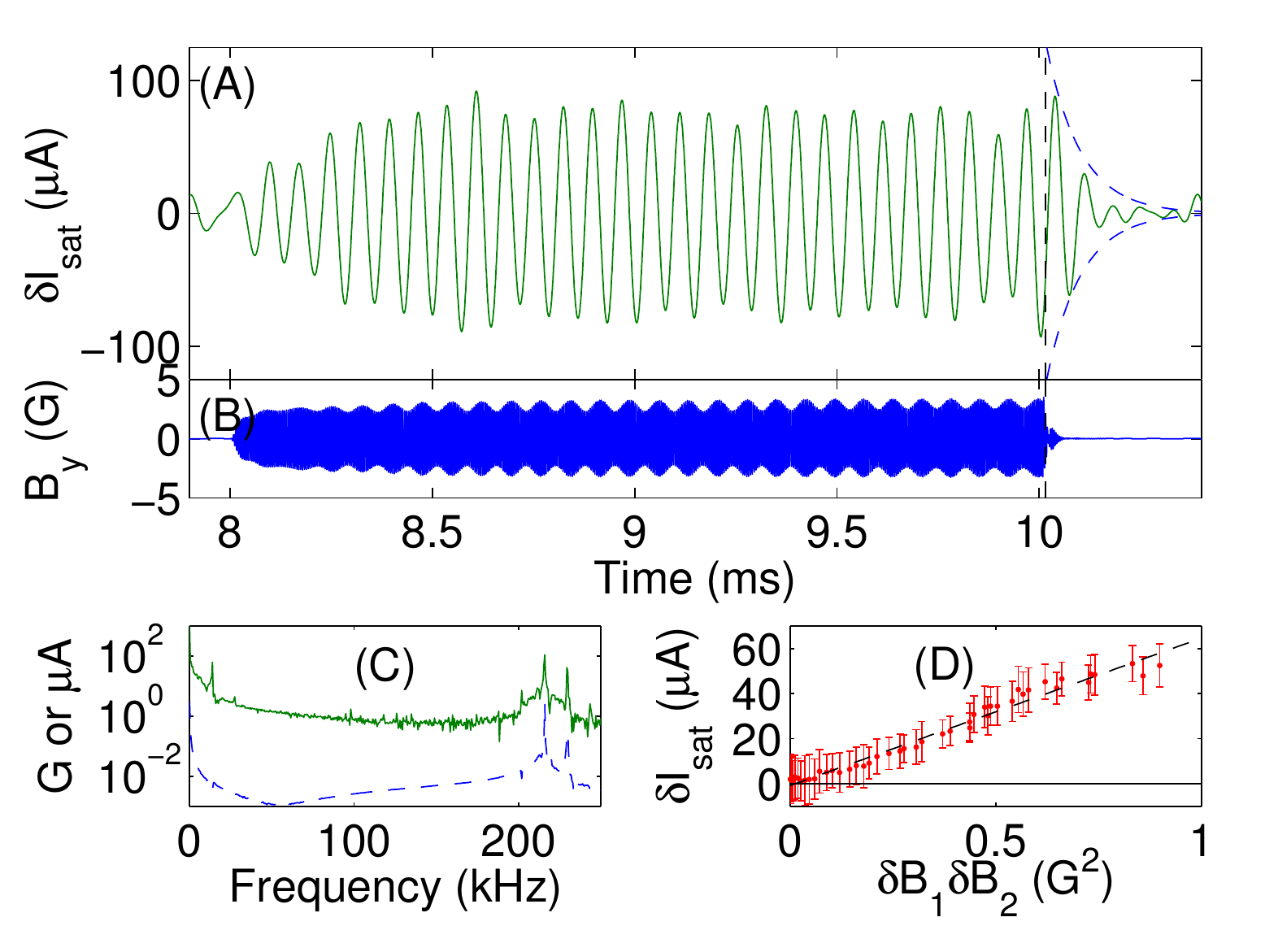}
\caption{Beat interaction between two counter-propagating Alfv{\'e}n waves in helium plasma showing the ring up and ring down associated with coupling to a natural mode of the plasma.  Two Alfv{\'e}n wave antennas at frequencies of $230$ and $216$~kHz are on between $8$ and $10$~ms.  $B_0=750$~G.  Panel~(A) shows the filtered ion saturation current signal between $8$~kHz and $20$~kHz from a Langmuir probe at $z=3.20$~m.  Panel~(B) shows a singal from a nearby magnetic probe at $z=2.24$~m.  The frequency spectrum over the $2$~ms window the antenna is on is displayed in Panel~(C) for the ion saturation current signal (upper solid trace) and the magnetic singal (lower dashed trace).  Panel~(D) shows the beat amplitude as a function of the product of the magnetic field amplitudes at the driving frequencies.  All probes are at $x=0$, $y=0$.}\label{fig:beatraw}
\end{figure}

A clear nonlinear response at the beat frequency is observed in these experiments, as shown in Fig.~\ref{fig:beatraw}.  When the two Alfv{\'e}n wave antennas are turned on between $8$~ms and $10$~ms in this helium discharge, a beat wave at the difference frequency of $14$~kHz is observed both in the filtered ion saturation current trace displayed in Panel~(A) and the full frequency spectrum shown in Panel~(C).  This signal will be shown to have many properties consistent with an ion acoustic mode produced by a three-wave matching process.  The beat amplitude of $75$~$\mu$A represents $\sim3.5\%$ of the measured mean ion saturation current.  After the last of the magnetic signatures from the Alfv{\'e}n waves pass by a fixed magnetic probe at $t=10.015$~ms, the amplitude of the beat wave does not immediately drop to zero, indicating that coupling to a normal mode of the plasma has occurred.  The ring-down time of the driven wave is $\sim 85$~$\mu${s}, comparable to an ion-neutral collision time of $\sim100$~$\mu$s for these parameters.  When these experiments are repeated in hydrogen plasmas, the ring down time is shorter; consistent with this, the ion-neutral collision frequency for the chosen parameters is larger in hydrogen than it is in helium \citep{janev87}.

\begin{figure}[tbp]
\centering
\includegraphics[width=\columnwidth]{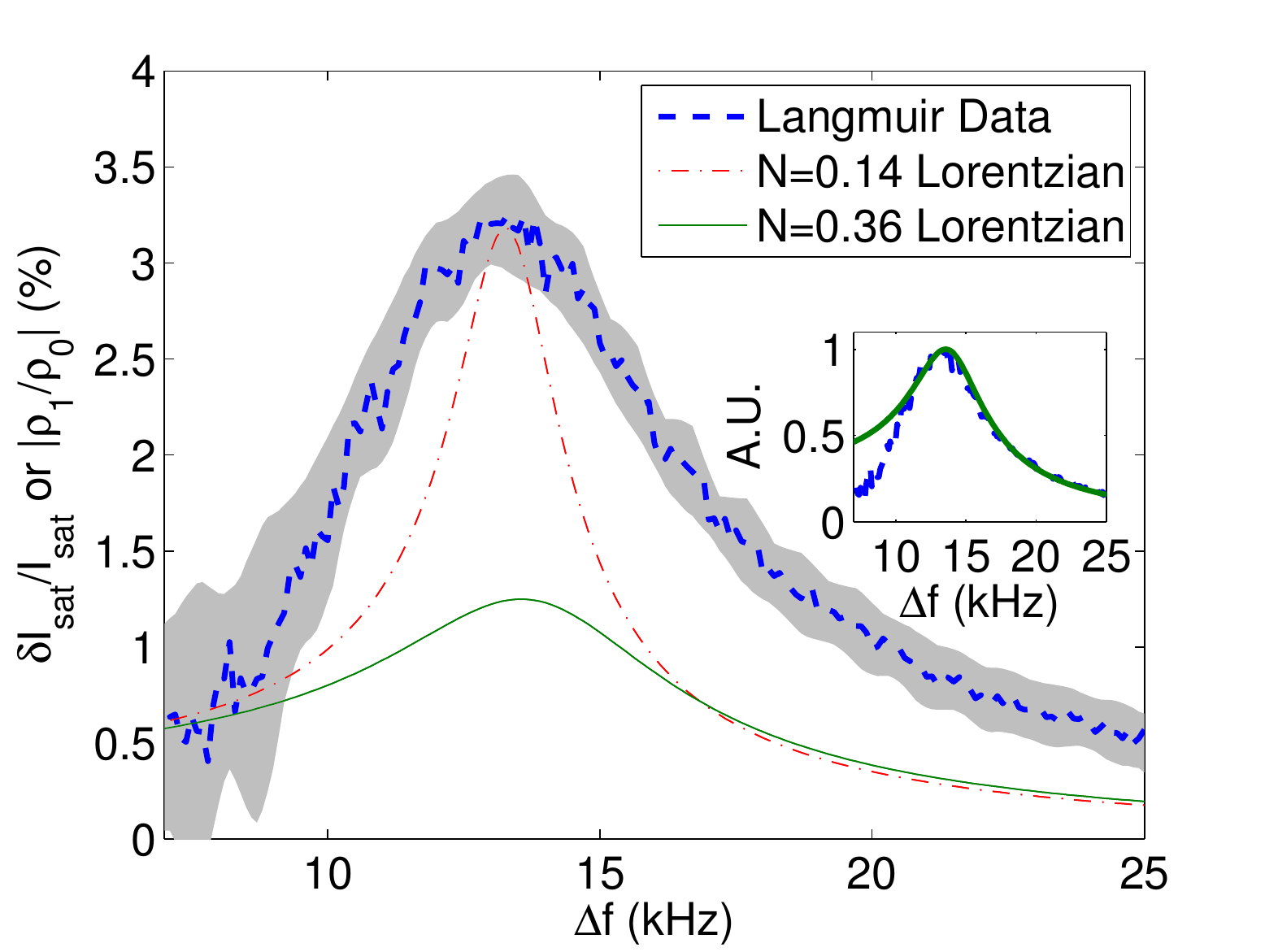}
\caption{Beat amplitude as a function of beat frequency $\Delta{f}$ showing a resonant response in helium plasma with background $B_0=750$~G.  The Alfv{\'e}n wave antenna on the cathode end produces a fixed frequency $230$~kHz wave while the frequency of the wave produced by the end mesh antenna is scanned from $205$ to $230$~kHz between discharges.  The dashed curve shows the beat amplitude $\delta{I_{sat}}$ as a percent of $I_{sat}$ observed as a function of $\Delta{f}$ for a Langmuir probe at $z=6.07$~m.  The amplitude is normalized to the zero frequency component; the shaded errorbar represents the level of background fluctuations.  The thin dash-dot trace is the predicted density response $|\rho_1/\rho_0|$ based on Eq.~\ref{eqn:respf} using $N = 0.14$,  $\beta = 4 \times 10^{-4}$ and $b_{\perp}/B_0 = {1/750}$.  The thin solid trace is the equivlent prediction for $N=0.36$.  The figure inset to the right shows the $N=0.36$ trace rescaled to the data to emphasise that this $N$ is the best fit for the width of the peak.}\label{fig:beatfscan}
\end{figure}

The beat amplitude is expected to be largest when three-wave coupling most efficiently excites a normal mode of the plasma.  The experimental strategy to test this prediction is as follows: the launch frequency of the cathode side antenna is held fixed while the launch frequency of the end mesh side antenna is varied between discharges.  The plasma response at the beat frequency is then examined in each discharge to find the difference frequency that best couples to an acoustic beat mode.  The results of this scan for helium plasmas with $B_0=750$~G are represented by the dashed line in Fig.~\ref{fig:beatfscan}.  This curve, representing the beat wave amplitude with both antennas on plotted as a function of the difference frequency, peaks at a frequency of around $13$~kHz.  This frequency at which three-wave matching relations are best satisfied to excite a normal mode of the plasma is defined as the resonance frequency.

A calculation based on the ion acoustic and Alfv{\'e}n wave dispersion
relations allows for a prediction of the observed resonance frequency.  For the measured
experimental parameters, $V_A/v_{\rm th,e} \lesssim 1$, suggesting
that a kinetic calculation of the dispersion relation would be
appropriate.  However, because the collisionality is fairly high
($\lambda_{\rm mfp,e} \sim 0.2$~m, $k_\parallel \lambda_{\rm mfp,e} < 1$) and in order to keep the
calculation simple, a fluid
dispersion relation is used for kinetic Alfv\'{e}n
waves~\citep{leneman99}:
$\omega=k_{||}V_A\sqrt{1+{\left(k_{\perp}\rho_s\right)^2}-\left({\omega
    / \Omega_i}\right)^2}$.  
The relevant dispersion relation for the ion acoustic mode is $\Delta\omega=\Delta{k_{||}}C_s/\sqrt{1+\left(\Delta{k_\perp}\rho_s\right)^2}$.  Three-wave matching relations predict that $\Delta{\omega}=\omega_2-\omega_1$, $\Delta{k_{||}}=k_{||2}+k_{||1}$, and $\Delta{k_{\perp}}=k_{\perp{2}}-k_{\perp{1}}$ where $1$ and $2$ are subscripts associated with the counter-propagating Alfv{\'e}n waves.  Some simple algebra and the assumption $\omega_1\approx\omega_2\equiv\omega\gg\Delta\omega$ lead to the equation:

\begin{equation}\label{eqn:beatf}
\Delta{\omega}={2\omega\displaystyle\sqrt{\beta} \over \sqrt{ 1+{\left(k_{\perp}\rho_s\right)^2}-\left(\displaystyle{\omega \over \Omega_i}\right)^2}}
\end{equation}

\noindent  Plugging in the experimental parameters used to produce Fig.~\ref{fig:beatfscan}, including a typical $\beta \sim 4 \times 10^{-4}$, Eq.~\ref{eqn:beatf} predicts a resonant frequency of $13$~kHz.  This agrees well with the experimental result.

\begin{figure}[tbp]
\centering
\includegraphics[width=\columnwidth]{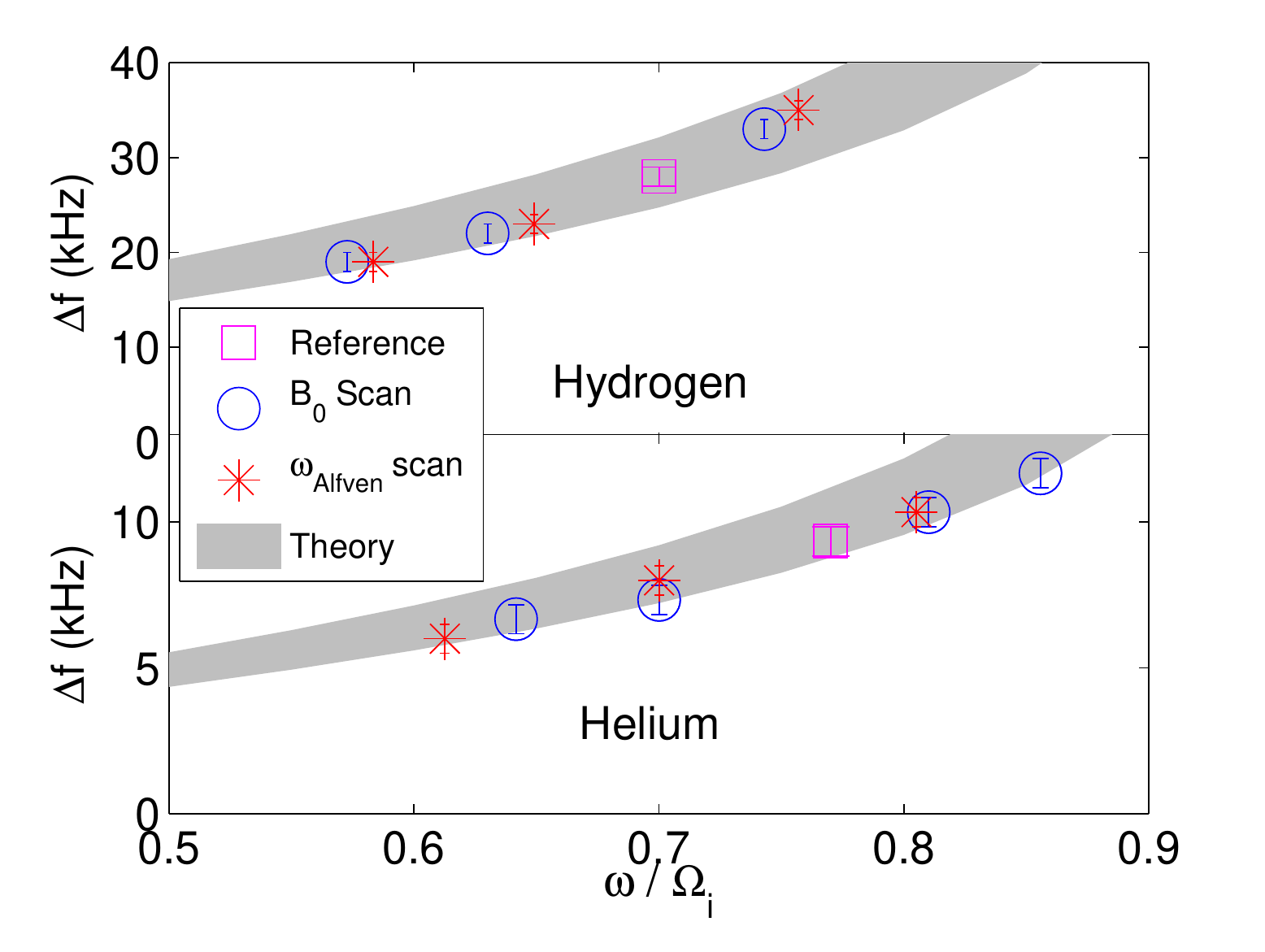}
\caption{Resonance frequency $\Delta{f}$ as a function of plasma and antenna parameters in both hydrogen and helium plasmas.  Data is for a Langmuir probe at $z=4.47$~cm in helium and $z=3.83$~cm in hydrogen.  The reference point is $\omega=480$~kHz, $B_0=450$~G in hydrogen and $\omega=220$~kHz, $B_0=750$~G in helium.  $k_\perp\rho_s$ is esimated from a Bessel function fit as $0.26$ in hydrogen and $0.28$ in helium.  Both magnetic field scans (circles) and antenna frequency scans (stars) are shown.  The gray shaded region represents the value of $\Delta{f}$ predicted by Eq.~\ref{eqn:beatf}, taking into account uncertainties in the density and temperature measured by the Langmuir probe.}\label{fig:beatf}
\end{figure}

Eq.~\ref{eqn:beatf} is satisfied for a wide range of plasma and antenna parameters; this is shown in Fig.~\ref{fig:beatf}.  For fixed ion mass, Eq.~\ref{eqn:beatf} implies that the resonant frequency is a function of $\omega/\Omega_i$.  Therefore, magnetic field scans and scans of the main antenna frequency may be overplotted on Fig.~\ref{fig:beatf}.  The resulting datapoints for each gas fall within the gray shaded region calculated using Eq.~\ref{eqn:beatf}.  The finite width of the gray region represents the statistical uncertainly from an average over similar Langmuir probe measurements.  Temperature is measured by sweeping the voltage of a single tip; density is obtained from the measured temperature and ion saturation current measurements.

To gain insight into the width of the response curve in Fig.~\ref{fig:beatfscan}, it is useful to model the non-linear interaction as a damped, driven oscillator.  In the simplest possible model, the ion acoustic perturbation is considered to be much smaller than the background density such that $\rho_1/\rho_0\ll{1}$ and $\rho_1/\rho_0\ll\nu/\omega$.  Self-consistent with this approximation, the parallel ion velocity perturbation is much smaller than the phase speed of the driven mode by the same order.  Combining the MHD momentum and continuity equations and neglecting perpendicular propagation effects:

\begin{equation}\label{eqn:mhdeq}
{\partial^2 \rho \over \partial t^2}+\nu{\partial \rho \over \partial t}-{C_s}^2{\partial^2 \rho \over \partial z^2}={\partial^2 \over \partial z^2}\left[{b_{\perp{1}} \cdot b_{\perp{2}} \over 4\pi}\right]
\end{equation}

Eq.~\ref{eqn:mhdeq} describes a damped, driven oscillator system.  The first and third terms represent the wave equation for an ion acoustic mode, the second term describes damping due to ion-neutral collisions, and the fourth term is the non-linear ponderomotive drive that results from interaction between the two Alfv{\'e}n waves.  This acoustic mode drive term accelerates ions parallel to $B_0$ through a non-linear $\tilde{v}\times\tilde{B}$ force in the parallel ion momentum equation.  For the ordering assumptions used to derive Eq.~\ref{eqn:mhdeq} to hold, the amplitude of the Alfv{\'e}n wave drive at resonance must be small, $b_{\perp}/B_0\ll{N\sqrt{\beta}}$, where $N=\nu/\omega_0$ represents the collisionality normalized to the resonance frequency.  During the time both Alfv{\'e}n waves are turned on, the system will respond at the drive frequency $\omega_D=\omega_2-\omega_1$ and the drive wave number $k_D=k_{||2}+k_{||1}$.  The response function at $\omega=\omega_D$ and $k=k_D$ follows from the linearization of Eq.~\ref{eqn:mhdeq}:

\begin{equation}\label{eqn:respf}
\left|{\rho_1 \over \rho_0}\right|={1 \over \beta}\left|{b_{\perp{1}} \over B_0}\cdot{b_{\perp{2}} \over B_0}\right|{1 \over \sqrt{\left(1-{\Omega_D}^2\right)^2+N^2{\Omega_D}^2}}
\end{equation}

\noindent where $\Omega_D=\omega_D/\omega_0$ is the drive frequency normalized to the natural resonance of the system $\omega_0=k_DC_s$.  Eq.~\ref{eqn:respf} is overplotted on Fig.~\ref{fig:beatfscan} for $N=0.14$ (thin dash-dot line) and $N=0.36$ (thin solid line).  While the amplitude of the resonant response is well predicted by $N=0.14$, the scaled inset figure shows that the best fit for the width of the peak is obtained for $N=0.36$.  Furthermore, $N=0.36$ is more than double the value $N=0.14$ obtained from the ringdown time in Fig.~\ref{fig:beatraw}.  This discrepancy is still under investigation and may be due to effects not included in the simple model, including finite perpendicular wave number and axial variation of plasma parameters.  Despite this, the $N=0.14$ result implies that the pondermotive force is of sufficient amplitude to drive the observed resonant response.  

\begin{figure}[tbp]
\centering
\includegraphics[width=\columnwidth]{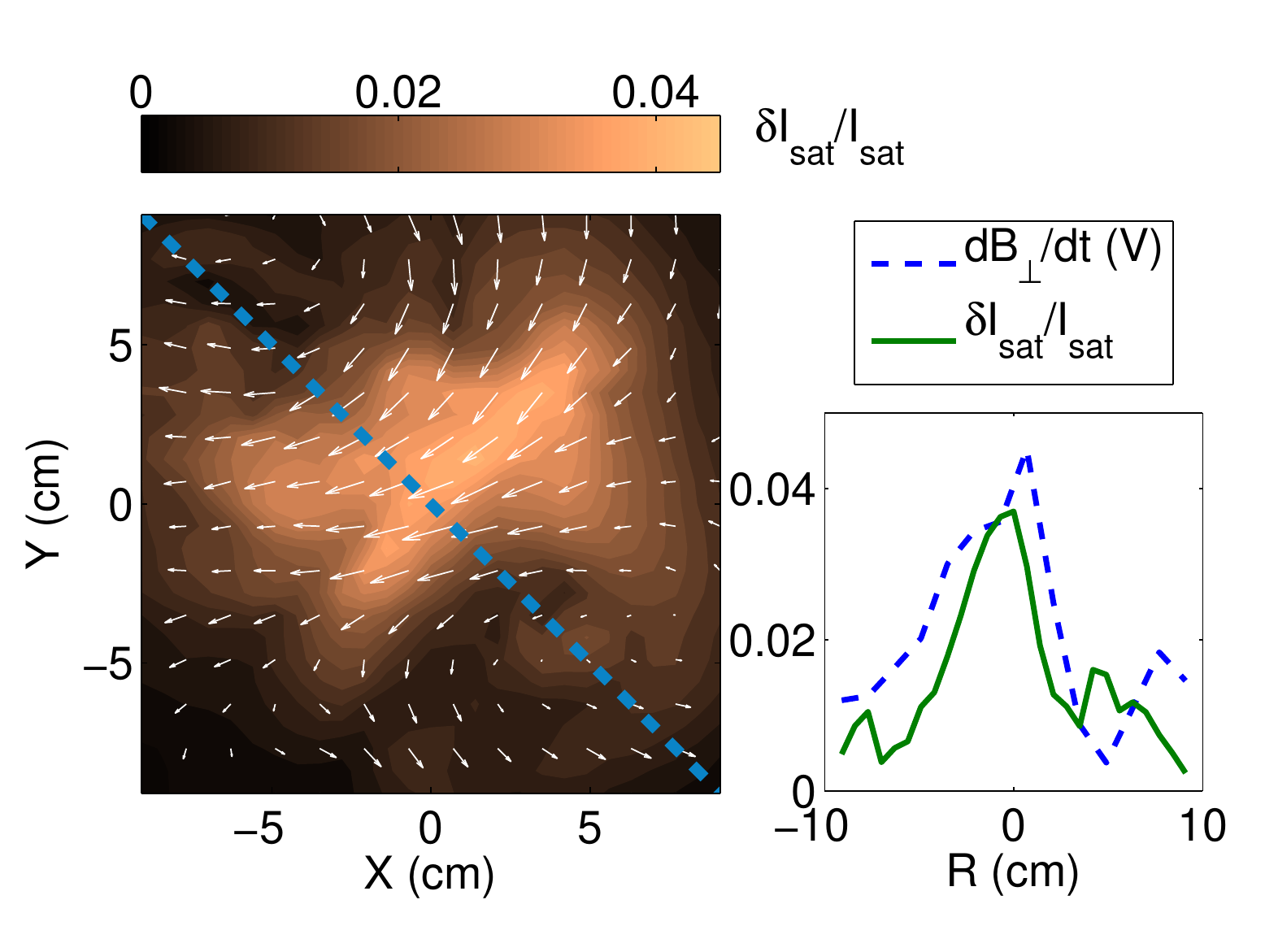}
\caption{Spatial profile of the beat and Alfv{\'e}n waves in hydrogen plasma.  On the left plot, white vectors represent the superimposed magnetic fields from the two Alfv{\'e}n waves while the color scale shows spatial variations of the beat amplitude.  The cathode and end mesh side antennas are set to $480$~kHz and $450$~kHz respectively.  $B_0=450$~G.  Magnetic data is averaged over the times at which the antenna-launched waves are in phase such that the magnetic field amplitude is at a maximum.  The two current channels of the Alfv{\'e}n waves are visible in the upper right and lower left.  In the right plot, the measured magnetic field amplitude and beat response are shown as a function of $R$, the distance from the orgin along the diagonal cut shown.  The beat amplitude is peaked near the origin where the local magnetic field peaks.}\label{fig:bprofile}
\end{figure}

The spatial profile of the beat response also suggests a pondermotive drive mechanism.  This is shown experimentally in Fig.~\ref{fig:bprofile}.  The measured wave magnetic field vectors are plotted as white arrows; overlapping current channels for the two Alfv{\'e}n waves are indicated by the circulation pattern of these arrows in the upper left and lower right portions of the figure.  The perpendicular wave number may be estimated by fitting the transverse magnetic field pattern of a single current channel to a spherical Bessel function of the first kind \citep{brugman07}; for the plasma and antenna parameters associated with Fig.~\ref{fig:bprofile}, this gives $k_\perp\sim 0.5$~/cm and $k_\perp\rho_s\sim 0.26$.  As indicated by the color scale, the beat amplitude is greatest near the origin which is where the Alfv{\'e}n wave magnetic field peaks.  Qualitatively, this result agrees well with Eq.~\ref{eqn:respf}.  Also consistent with the MHD theory presented, a scan of the antenna power (Fig.~\ref{fig:beatraw}, Panel~(D)) reveals that the beat-driven amplitude grows proportionally to the product of the two Alfv\'{e}n wave amplitudes.

\begin{figure}[tbp]
\centering
\includegraphics[width=\columnwidth]{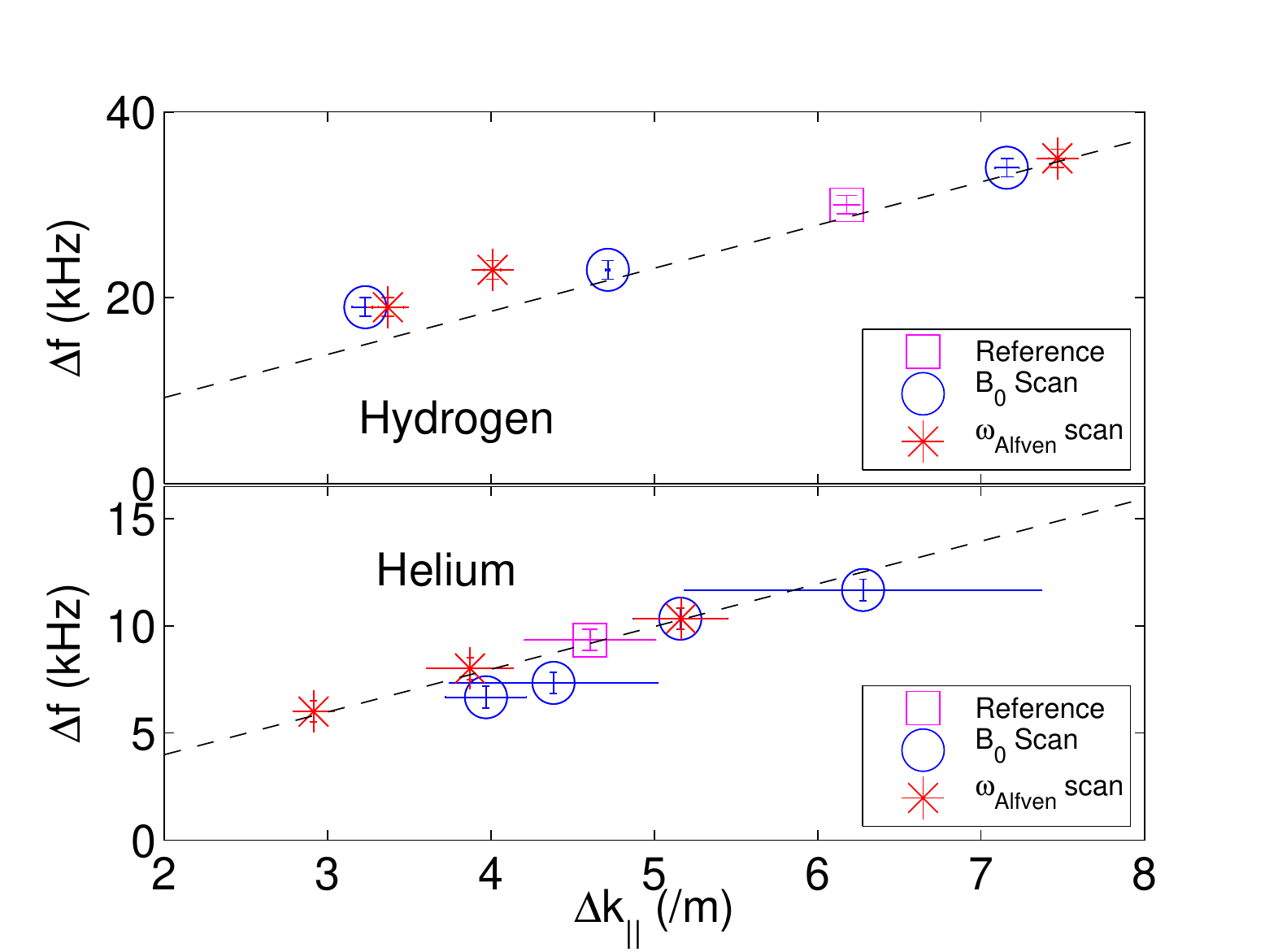}
\caption{Dispersion relation of beat waves at the resonance frequency.  Experimental runs are the same as in Fig.~\ref{fig:beatf}.  The best fit dashed line represents a phase speed of $29.1\pm0.7$~km/s in the hydrogen plot and $12.5\pm0.3$~km/s in the helium plot.}\label{fig:disp}
\end{figure}

The resonance in the beat wave response is identified as an ion acoustic mode based on the dispersion relation.  For each of the experimental runs in Fig.~\ref{fig:beatf}, the parallel wave number for the response at the resonance frequency is determined by examining the phase delay  between two Langmuir probes closely spaced in $z$ (One such experimental setup is shown in Fig.~\ref{fig:esetup}).  The result, shown in Fig.~\ref{fig:disp}, is a linear dispersion relation with phase speed comparable to the sound speed for LAPD parameters.  Using a kinetic dispersion relation \citep{hasegawa76} for ion acoustic modes and assuming an ion temperature of $1$~eV (previously measured in both helium and argon \citep{palmer05}), the phase speed in helium requires $T_e=4.4\pm0.3$~eV, well in line with the value of $4.3\pm0.9$~eV obtained by analyzing Langmuir probe sweeps during the beat wave time period.  Applying the same set of assumptions to the hydrogen data requires $T_e=6.7\pm0.4$~eV, well above the value of $4.3\pm1.0$~eV measured.  Since ion temperature has not been measured in hydrogen, one possible explanation for this discrepancy is $T_i>1$~eV.  This could be a combination of higher background ion temperature and enhanced ion heating by the launched Alfv{\'e}n waves in hydrogen plasmas, due to the lighter ion mass.  For example, if the ion temperature is $2$~eV, the ion acoustic mode dispersion relation \citep{hasegawa76} requires only $T_e=5.2\pm0.3$~eV which is within errorbar of the measured value.

In summary, the first laboratory observations of the Alfv{\'e}n-acoustic mode coupling at the heart of parametric decay are presented.  Counter-propagating Alfv{\'e}n waves launched from either end of the LAPD produce a resonant response identified as an ion acoustic mode based on the dispersion relation, spatial profile, and other features consistent with a simple MHD theory.  Several areas for further investigation remain.  Ion acoustic waves have never been directly launched by an antenna in the laboratory at densities comparable to those in the LAPD.  Thus, a new technique to directly launch ion acoustic waves is being developed and will be utilized for a detailed study of the damping mechanism.  The new technique will also be used to investigate Alfv{\'e}n and ion acoustic wave coupling; the launched acoustic mode could potentially seed the parametric decay process.  Additional studies may also focus on parametric decay from a single large amplitude Alfv{\'e}n wave; this is not possible under the present set of parameters due to insufficient Alfv{\'e}n wave amplitude.

\begin{acknowledgments}
The authors thank G.~Morales and J.~Maggs for insightful discussions, P.~Pribyl, S.~K.~P.~Tripathi, B.~Van Compernolle, and S.~Vincena for insightful discussions and assistance with the experiment and Z.~Lucky and M.~Drandell for their excellent technical support.  S.~D.~was supported by a DOE FES Postdoctoral Fellowship.  This work was performed at the UCLA Basic Plasma Science Facility which is supported by DOE and NSF.
\end{acknowledgments}


\end{document}